\documentclass [notitlepage]{article}

\begin{document}

\title{A comment on the two definitions of the electric polarizability of a bound system in relativistic quantum theory}
\author{M.A. Maize\\Department of Physics\\Saint Vincent College\\300 Fraser Purchase Rd\\Latrobe, PA 15650}
\date{}
\maketitle

In 1999, F.A.B. Coutinho, Y.Nogami and Lauro Tomio published a paper entitled ``Two definitions of the electric polarizability of a bound system in relativistic quantum theory''$^1$.  The authors of ref\@.(1), dicussed the possibility of two definitions of the relativistic electric polarizability and the difference between the two definitions$^1$.  In addition, the authors relate their definitions to the electric polarizability as it appears in a limited selection of studies.  The authors of ref\@.(1) made serious mistakes in their analyses and conclusion.  In addition, they ignored very valuable references which are essential to the problem of relativistic electric polarizability.  It is the purpose of this note to point out the essential mistakes of ref\@.(1) and to point to some of the references which are of fundamental value to studying high energy photon scattering and the relativistic electric polarizability.

The two definitions of the relativistic electric polarizability in ref\@.(1) are $\alpha$ and $\alpha_{QM}$.  $\alpha$ is introduced as the relativistic electric polarizability which is obtained in quantum field theory while $\alpha_{QM}$ is introduced as the relativistic electric polarizability which is obtained in single-particle quantum mechanics$^1$.  The authors of ref\@.(1) stress that when the vacuum background is included in the system, $\alpha$ is obtained while $\alpha_{QM}$ is obtained when the vacuum background is not included in the system$^1$.

The electric polarizabilities in ref\@.(1) are defined as$^1$: \begin{equation} W_{QM} = -1/2 \alpha_{QM} \mbox{\boldmath $E^2$} \ \ \ ,\ \ \ W_{QFT} = -1/2 \alpha \mbox{\boldmath $E^2$} \ \ \ \ , \end{equation} where \mbox{\boldmath $E$} is the external electric field while $W_{QM}$ and $W_{QFT}$ represent the second-order energy shift in single-particle quantum mechanics and quantum field theory respectively.  $W_{QM}$ is given by$^1$: \begin{equation} W_{QM} = \sum_{i\neq1} \frac{|V_{i,1}|^2}{E_1 - E_i} + \sum_{j} \frac{|V_{-j,1}|^2}{E_1 - E_{-j}} \ \ \ \ ,\end{equation} where V is the external perturbation and is equal to -q\mbox{\boldmath $E.r$} with q being the charge of the particle which occupies the state $|1>$.  The states $|i>$'s and $|-j>$'s are the positive and negative energy states respectively$^1$.  Also the $|i>$'s and $|-j>$'s are the solution of the unperturbed hamiltonian which includes the binding potential$^1$.  The state $|1>$ is the lowest position energy state.  \mbox{$V_{i,1} = <i|V|1>$} and \mbox{$V_{-j,1} = <-j|V|1>^1$.}  The expression for $W_{QFT}$ is given by$^1$: \begin{equation} W_{QFT} = \sum_{i\neq1} \frac{|V_{i,1}|^2}{E_1 - E_i} + \sum_{i\neq1,j} \frac{|V_{i,-j}|^2}{E_{-j} - E_i} \ \ \ \ . \end{equation}  With $W'_{vac}$ defined $as^1$:  \begin{equation} W'_{vac} = \sum_{i,j} \frac{|V_{i,-j}|^2}{E_{-j}-E_i} \ \ \ \ , \end{equation} the authors of ref\@.(1), conclude the following: \begin{equation} W_{QFT} = W_{QM} + W'_{vac} \ \ \ \ , \end{equation} and \begin{equation} \alpha = \alpha_{QM} + \alpha'_{vac} \ \ \ \ , \end{equation} where $\alpha'_{vac}$ is the electric polarizability of the vacuum in the absence of the particle in $|1>^1$.  Then, the authors of ref\@.(1) use the hydrogen atom as an example.  They define $\alpha$ as the electric polarizability of the atom$^1$.  They define $\alpha'_{vac}$ as the electric polarizability of the hydrogren ion$^1$.  No definition was given to $\alpha_{QM}$ in terms of the hydrogen atom in ref\@.(1).  At this point some remarks are in order.

The authors of ref\@.(1), cite the work by Sucher$^2$ to relate their definitions to his definition.  In reference to Sucher$^2$, they write the following$^1$: \begin{quote} He says that the EP defined in terms of second-order perturbation theory always gives a positive value (negative energy shift) for a system in its ground state.  \end{quote}Then they add$^1$: \begin{quote} The EP that he refers to is, in our notation, $\alpha$ and not $\alpha_{QM}$. \end{quote}  Sucher writes the following in his paper$^2$: \begin{quote} Within the framework of nonrelativistic quantum mechanics the positivity of $\alpha_E$ is easily understood.  The second-order level shift W induced by the perturbation $H' = -\sum_{i} e_i \mbox{\boldmath $r_i.E$}$, where the $e_i$ are the constituent charges and \mbox{\boldmath $E$} is a uniform electric field, has the form $W = -1/2 \alpha_E\mbox{\boldmath $E^2$}$, which defines $\alpha_E$, and second-order perturbation theory always gives a negative energy shift for a system in its ground state.  \end{quote}  The authors of ref\@.(1) define $\alpha$ as a relativistic electric polarizability and then set it equal to the nonrelativistic electric polarizability.  In addition, the authors of ref\@.(1) discuss the assumption of the high energy limit which was set by Sucher$^2$.  Sucher emphasized that a high energy assumption is necessary to derive$^2$ $\alpha_E$(electric polarizability) $> 0$.  Sucher's high energy assumption$^2$ is that $f_E(\omega)$ (the forward scattering amplitude) goes to zero when $\omega$ (the energy) goes to $\infty$.  $f_E(\omega)$ and the static polarizability are related by the equation$^2$ $f_E(0) = 4\pi\alpha_E$.  The authors of ref\@.(1), conclude that when vacuum background is included $\alpha$ is obtained.  They also conclude that since $\alpha$ is always positive, it is the same as $\alpha_E$.  However, the authors of ref\@.(1) never explained why including the vacuum background is equivalent to setting the high energy limit regarding the scattering amplitude$^2$.

In 1959, Erber published a paper in which he discussed the coherent Compton amplitude at high energies$^3$.  In his paper, Erber considered the high energy limit of the forward scattering amplitude and the scattering cross section in connection with the application of dispersion relations.  One of Erber's results is that $\sigma(bound electron) = \sigma(atom) - \sigma(proton)$, where $\sigma$ refers to the scattering cross section.  The author's of ref\@.(1), never referred to the bound electron when they gave their example of the hydrogen atom.  This is expected, since they ignored in their example the relationship between the scattering cross section and the electric polarizability.  In addition $\sigma(bound electron)$ is then related to D(bound electron) in the dispersion relation used by Erber$^3$.  $D(bound electron) = D(atom) - D(proton)$ where D refers to the real part of the forward scattering amplitude.  The subtraction existing in the dispersion relation$^3$ and its effect on calculating the electric polarizability was completely ignored in ref\@.(1).

In 1968, Goldberger and Low published a paper entitled ``Photon scattering from bound atomic systems at very high energy''$^4$.  In their paper$^4$, the bound-electron scattering is defined as the scattering by the atom minus the scattering by the nuclear Coulomb field in the absence of the electron.  Goldberger and Low also add the following: \begin{quote}The contribution from pair production by the atom with the electron going to any state other than the inital bound state is precisely cancelled by the pair production in the absence of the inital electron, as long as electron interactions are neglected.$^4$.\end{quote}  As stressed in refs\@.(3) and (4), the scattering by the bound electron in the relativistic case should be properly identified.  This point certainly affects the calculation of the relativistic electric polarizability of a bound system.  Refs\@.(3) and (4) in addition to any discussion related to the scattering by the bound electrons were ignored in ref\@.(1).  This is puzzling since the hydrogen atom was used as an example of ref\@.(1).  In addition the pair terms and their contribution to the scattering cross section and the electric polarizability were never mentioned explicitly or discussed in ref\@.(1). 

Finally, the authors of ref\@.(1) say that we suggested in our work$^5$, that we are giving the reasons to why the electric polarizability is negative based on Sucher's definition.  We cited the paper of Sucher$^2$, in our work $^5$ as an example of the possibility of negative electric polarizability.  As we stated in ref\@.(5), we used our model to find the conditions which lead to negative electric polarizability.  As we stressed in ref\@.(5), our definition of the relativistic electric polarizability is constistent with the discussion and the definitions of refs\@.(3) and (4).  Such a fact was ignored in ref\@.(1) as well as ignoring very valuable references such as refs\@.(3) and (4).  

Finally, we would like to stress that subtraction in the dispersion relation which was ignored in ref\@.(1), had been studied a long time ago$^6$.  In addition, model calculations producing negative electric polarizability have been in existance for years$^{7,8}$.  The article by \mbox{J.L. Friar$^9$} would be a good example of a review paper discussing the subject of polarizability, Compton scattering and relevant dispersion relations in addition to other topics.

\end{document}